\documentclass[aps,twocolumn, superscriptaddress]{revtex4-2}
\usepackage{txfonts}
\usepackage{graphicx}
\usepackage{dcolumn}
\usepackage[T1]{fontenc} 
\usepackage{subfig}
\usepackage{color}
\usepackage{hyperref}
\hypersetup{
    colorlinks=true,
    linkcolor=blue,
    filecolor=blue,      
    urlcolor=blue,
    citecolor=blue,
    pdfpagemode=FullScreen,
    }
\usepackage{lipsum}
\usepackage{soul} 
\usepackage{svg}
\usepackage{xr}
\usepackage{verbatim} 

\def\updn{\uparrow\downarrow}
\def\dnup{\downarrow\uparrow}

\draft 
\DeclareGraphicsExtensions{.png,.svg,.jpeg}
\begin{document}

\title{Coexistence of Anomalous Hall Effect and Weak Net Magnetization in
Collinear Antiferromagnet MnTe} 

\author{K. P. Kluczyk$^\ddag$}
\affiliation{Faculty of Physics, University of Warsaw, Pasteura 5, Warsaw, Poland}
\thanks{These two authors contributed equally.}

\author{K.~Gas$^\ddag$}
\affiliation{Institute of Physics, Polish Academy of Sciences, Aleja Lotnik\'ow 32/46, PL-02668 Warsaw, Poland}
\affiliation{Center for Science and Innovation in Spintronics, Tohoku University,
Sendai 980-8577, Japan}
\thanks{These two authors contributed equally.}

\author{M. J. Grzybowski}
\affiliation{Faculty of Physics, University of Warsaw, Pasteura 5, Warsaw, Poland}

\author{P.~Skupi\'nski}
\affiliation{Institute of Physics, Polish Academy of Sciences, Aleja Lotnik\'ow 32/46, PL-02668 Warsaw, Poland}

\author{M.~A.~Borysiewicz}
\affiliation{Łukasiewicz Research Network - Institute of Microelectronics and Photonics, Aleja Lotnik\'ow 32/46, PL-02668 Warsaw, Poland}

\author{T.~Fąs}
\affiliation{Faculty of Physics, University of Warsaw, Pasteura 5, Warsaw, Poland}

\author{J.~Suf\mbox{}fczy\'nski}
\affiliation{Faculty of Physics, University of Warsaw, Pasteura 5, Warsaw, Poland}

\author{J.~Z.~Domagala}
\affiliation{Institute of Physics, Polish Academy of Sciences, Aleja Lotnik\'ow 32/46, PL-02668 Warsaw, Poland}

\author{K.~Grasza}
\affiliation{Institute of Physics, Polish Academy of Sciences, Aleja Lotnik\'ow 32/46, PL-02668 Warsaw, Poland}

\author{A.~Mycielski}
\affiliation{Institute of Physics, Polish Academy of Sciences, Aleja Lotnik\'ow 32/46, PL-02668 Warsaw, Poland}

\author{M.~Baj}
\affiliation{Faculty of Physics, University of Warsaw, Pasteura 5, Warsaw, Poland}

\author{K.~H.~Ahn}
\affiliation{FZU-Institute of Physics of the Czech Academy of Sciences, Cukrovarnick\'a 10, CZ-16253 Praha, Czech Republic}

\author{K.~V\'yborn\'y}
\affiliation{FZU-Institute of Physics of the Czech Academy of Sciences, Cukrovarnick\'a 10, CZ-16253 Praha, Czech Republic}

\author{M.~Sawicki}
\email[]{mikes@ifpan.edu.pl}
\affiliation{Institute of Physics, Polish Academy of Sciences, Aleja Lotnik\'ow 32/46, PL-02668 Warsaw, Poland}

\author{M.~Gryglas-Borysiewicz}
\email[]{mgryglas@fuw.edu.pl}
\affiliation{Faculty of Physics, University of Warsaw, Pasteura 5, Warsaw, Poland}

\date{\today}

\begin{abstract}
    Anomalous Hall effect (AHE) plays important role in the rapidly developing field of antiferromagnetic spintronics. It has been recently discussed that it can be a feature of not only uncompensated magnetic systems but also in altermagnetic materials. Hexagonal MnTe belongs to this appealing group of compounds exhibiting AHE and is commonly perceived as magnetically compensated. Here, we demonstrate that bulk form of MnTe exhibits small but detectable magnetic moment correlating with hysteretic behaviour of the AHE. We formulate a phenomenological model 
    which explains how this feature allows to create a disbalance between states with
    opposite N\'eel vector and prevent the AHE signal from averaging out to zero.
    Moreover, we show how the dependence of AHE on the N\'eel vector arises on 
    microscopical level and highlight the differences in Berry curvature between
    magnetically compensated and uncompensated systems. 
    
\end{abstract}


\maketitle

\section{Introduction}
    Antiferromagnetic materials attract a lot of attention due to the groundbreaking demonstrations of current-induced control of the spin axis \cite{Wadley2016Science, Moriyama2018SR, Baldrati2019PRL}, recently observed long-distance spin transport \cite{Lebrun2018Nature, Han2020} and numerous fundamental findings related to the role of the local atomic site asymmetries. They can give rise to the N\'eel order spin orbit torques \cite{Zelezny2014, Zhang2014} or recently recognized new group of magnetic materials called altermagnets \cite{Smejkal2022,Smejkal2022b} in which different spin sublattices are coupled by certain rotation symmetry operations giving rise to multiple interesting phenomena including large bands spin-splitting. The intensive interest in both ordinary antiferromagnets (AFs) and altermagnets results in rapid development of the array of experimental methods capable to detect the reorientation of the N\'eel vector \cite{Nemec2018, Gray2019PRX, Gross2017}. The electrical methods such as anisotropic magnetoresistance (AMR) \cite{Marti2014, Wang2020PRB, AMR-review}, spin Hall magnetoresistance (SMR) \cite{Hoogeboom2017APL, Baldrati2018PRB, Fischer2018PRB, Geprags2020} or anomalous Hall effect (AHE) give insight into the magnetic properties. Remarkably, AHE was shown to be present in systems that are magnetically compensated such as non-collinear antiferromagnets \cite{Nakatsuji2015, Nayak2016, Suzuki2016} or altermagnets \cite{Feng2022, Betancourt2023} including a very compelling, semiconducting hexagonal MnTe. The very fact that MnTe is conducting allows to directly probe its properties in electronic transport experiment.


    
    In this report, we have addressed high quality free-standing bulk samples eliminating the influence of {\color{black} epitaxial strain} on the magnetic anisotropy \cite{Chen2018PRL} and the domain structure \cite{Grzybowski2017PRL, Wadley2018NN} opening a way for {\color{black} precise quantitative volume magnetometry due to the absence of bulky substrates. {\color{black} Also, for bulk samples the impact of possible parasitic substrate or interface conductivity may be neglected}. 
    The magnetization and resistivity studies both with the Raman spectroscopy evidenced the antiferromagnet-paramagnet phase transition. Moreover, not only do we show that the anomalous Hall effect is present also in bulk samples, complementing the recent study of thin layers epitaxially grown on InP(111)A substrates~\cite{Betancourt2023}, but also we demonstrate the existence of relatively weak ferromagnetic-like (WFL) signal, which presence correlates with the AHE and the antiferromagnetic (AF) state. 
The anomalous Hall signal recently detected in MnTe~\cite{Betancourt2023}, is believed to originate from the interplay of non-relativistic \footnote{We use the term 'non-relativistic' here in the same sense as in~\cite{Smejkal2022b}: 
    while the magnetic structure of the system is assumed to be the same as observed experimentally, symmetries of the Hamiltonian are analysed as if it did not contain any spin-orbit interaction.
    The absence of PT and tT symmetries then allows to resolve band degeneracy (here T stands for time reversal and P, t for spatial symmetries: inversion or certain translation). Ensuing momentum-dependent spin splitting allows for efficient
    spin-charge conversion~\cite{Yuan:2020_a,RGH:2021_a}.
    }
    symmetry-related spin splitting of bands and spin-orbit interaction.
This can result in: (1) the dependence of AHE on the direction of the N\'eel vector and (2) the effective control of magnetic moments in the basal plane $\perp c$ by external magnetic field applied along the $c$-axis.
  
To substantiate this notion we formulate a phenomenological macrospin model of this material system. The model explains the existence of finite magnetic moment (even in the absence of an external magnetic field) by the interaction of the Dzyaloshinskii–Moriya type and for a reasonable set of parameters it can reproduce well the hysteretic behavior of 
AHE. While in real samples the finite net magnetization $\vec{M}_{net}$ is essential to achieve an uncompensated domain structure (where the AHE does not average out to zero), a hypothetical single-domain MnTe system can exhibit AHE even if the magnetic moments lie exactly in-plane (i.e. $\vec{M}_{net}=0$), as shown by Berry curvature (BC) analysis.

\section{Samples}
    
   Bulk MnTe crystals were produced from vapour phase. 5N~tellurium and manganese powders were placed at two ends of quartz ampule, which was subsequently sealed under vacuum and heated. After evaporation of tellurium at temperature of 600$^{\circ}$C, at about 950$^{\circ}$C tellurium vapors reacted with manganese, forming few mm large MnTe crystals of irregular shape. The composition of the crystals was studied with X-ray diffraction experiment which revealed the desired NiAs-type MnTe (Fig.~\ref{fig:CrystalStruct}) and no alien phases 
   {(see Sec.~S.~I. in Supplementary Materials~\cite{our-SI}}). Preliminary transport experiments were performed on samples of irregular shapes. For quantitative transport experiments a small bar structure was prepared (with the width and heights of about 0.2~mm, and length of about 2~mm). Two pairs of contacts in a six terminal Hall configuration (Fig.~\ref{fig:SampleConfiguration}) were made on each side of the bar by gold pre-deposition and silver-based electrically conductive epoxy. 

    \begin{figure}[h!]
        \centering
            \includegraphics[width=6.5cm]{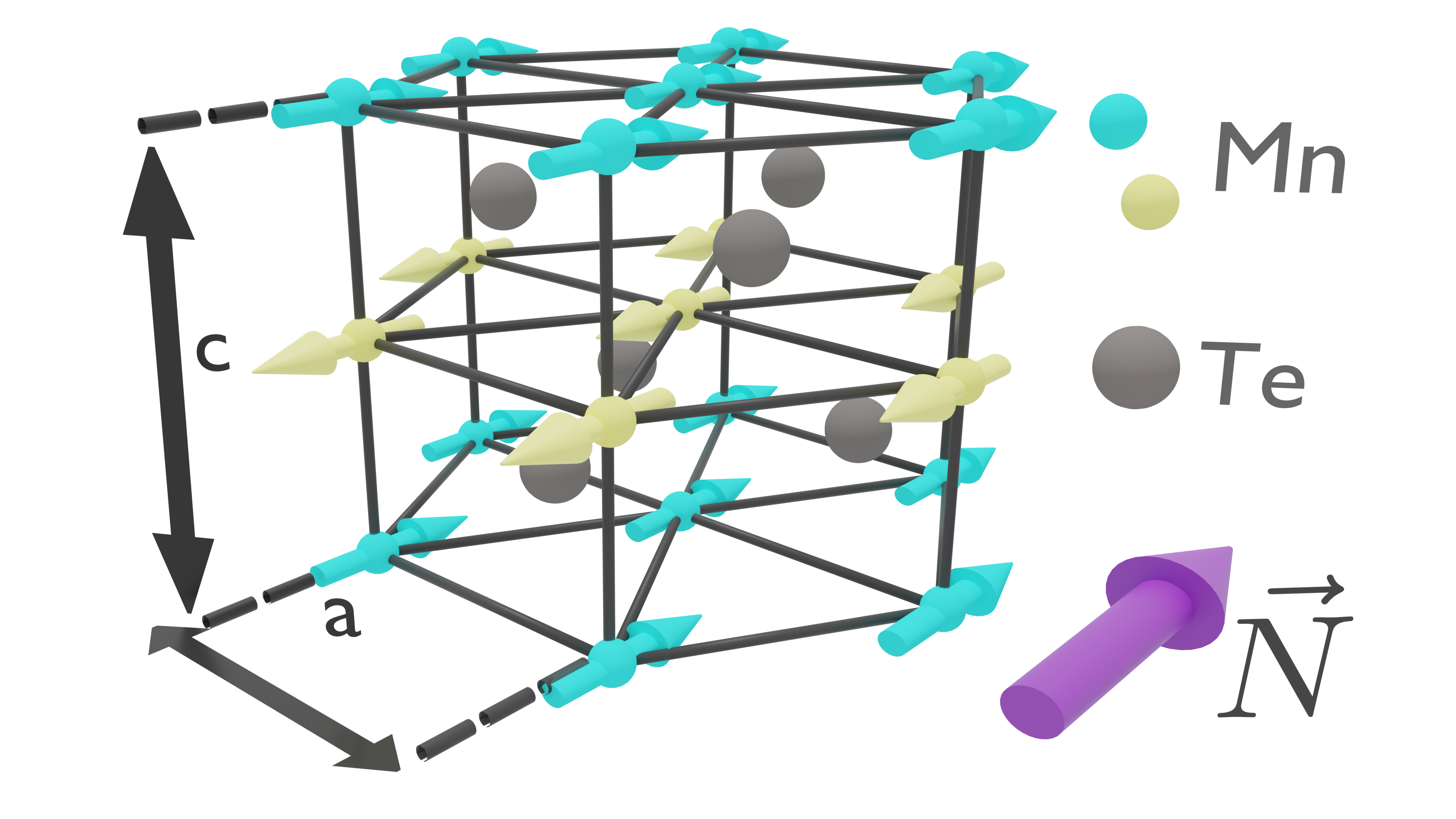}
            \caption{Crystal structure of hexagonal MnTe, with cyan and yellow balls representing Mn atoms with arrows indicating magnetic moments. Violet arrow shows N\'eel vector orientation. \label{fig:CrystalStruct}}
            \includegraphics[width=6.75cm]{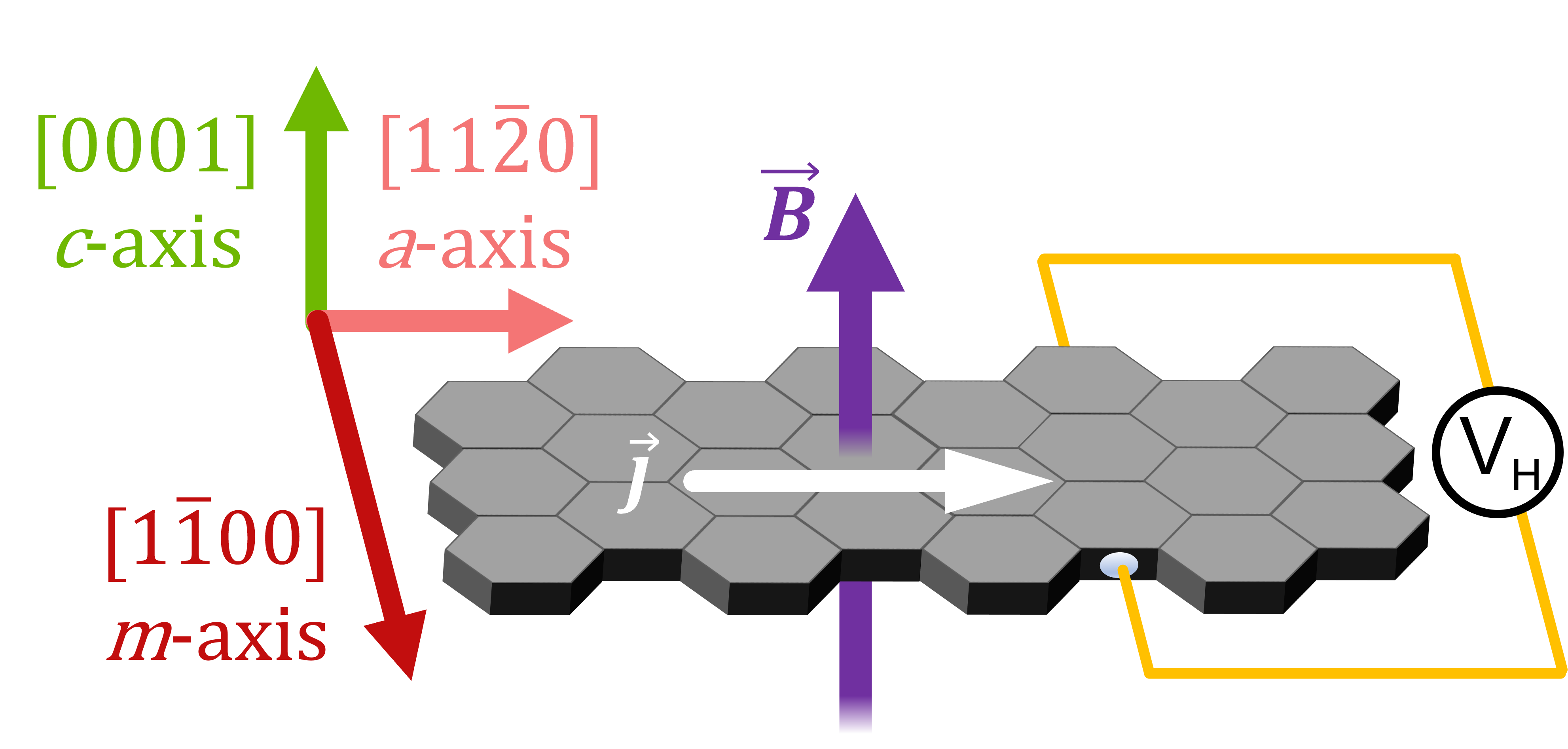}
            \caption{Configuration of the measuring system in relation to the crystal structure of the sample.\label{fig:SampleConfiguration}}
    \end{figure}

    Both orientation of this hallbar and its homogeneity were determined in high resolution X-ray diffraction experiment (HR~XRD), showing that $c$ axis was perpendicular to the current, which was along [11$\overline{2}$0] direction (see Fig.\ref{fig:SampleConfiguration} for sample scheme), with the accuracy of $5^{\circ}$. Besides, a formation of grains with a slight disorientation of $\sim0.20^{\circ}$ was also evidenced. The calculated lattice parameters are: $c=6.7116~\AA$, $a=4.1483~\AA$, as expected for the hexagonal MnTe and slightly different than for the MBE-grown samples \cite{Przezdziecka_2005, Kriegner2016}.    

    Additionally, we characterise our samples using Raman spectroscopy 
    ({see Sec IIIA of the Supplementary Information}). A dependence on temperature of spectral properties of $E_{2g}$ phonon transition (intensity, energy and linewidth) exhibits a clear peculiarity in the vicinity of the N\'eel temperature. {\color{black}This allows us to independently confirm the $T_{\text{N}}$ value.}

    

\section{Experimental Results}

\subsection{Magnetic Properties}
    Magnetic measurements were performed using superconducting quantum interference device (SQUID) magnetometer~MPMS~XL5 of Quantum Design. Two specimens were investigated. For temperature dependent studies at weak magnetic fields exactly the same sample was used as for the transport measurements, however after removing the electrical connections. Since some residues of silver epoxy remained on the specimen, for high field studies another, nearly cube-shaped sample was prepared of sub-millimeter dimensions, cut from a neighboring location to the transport sample, with crystallographic orientation confirmed by HRXRD. For the magnetic studies the samples are fixed to the sample holders formed from about 20 cm long and 1.7 mm wide Si sticks by means of strongly diluted GE varnish. We strictly follow all rules formulated for precision magnetometry and quantitative determination of weak magnetic signals in commercial magnetometers~\cite{Sawicki_2011}. 

    \begin{figure}[h!]        
              \includegraphics[width=8.2cm]{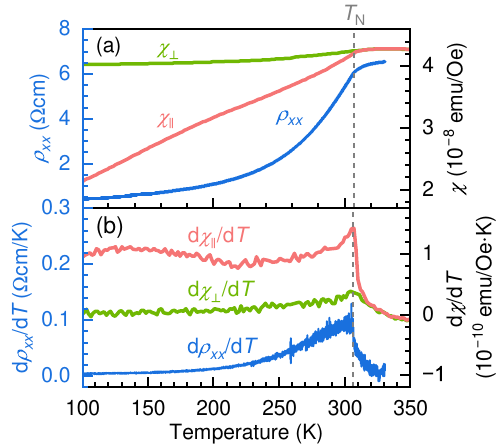}
            \caption{(a) Resistivity and magnetic susceptibility dependence on temperature of the studied hexagonal MnTe samples.
            (b) The peak in the first derivative of susceptibility $\mathrm{d}\chi/\mathrm{d}T$ and resistivity $\mathrm{d}\rho_{xx}/\mathrm{d}T$ at the N\'eel temperature. \label{fig:chirhoT}}
    \end{figure} 

    The samples do exhibit typical basic magnetic characteristics expected for highly oriented hexagonal AF crystals. Namely, a sizable magnetic anisotropy \cite{Coey} with respect to the hexagonal $c$-axis was observed, as presented in Fig.~\ref{fig:chirhoT}(a). At low temperatures the magnetic susceptibility,~$\chi$, measured along the $c$-axis $\chi_{\parallel}$ is substantially smaller than that measured with magnetic field $H$ perpendicular to $c$-axis,~$\chi_{\perp}$,
    {\color{black} revealing the in-plane orientation of the N\'eel vector.}
    Both $\chi_{\parallel}$ and $\chi_{\perp}$ coincide just above 300~K and follow same trend on further warming. Those results stay in perfect agreement with results of Komatsubara et al.~\cite{Komatsubara}. Despite the fact that both dependencies pass through a weak maximum at about 330~K we assign the temperature the two curves merge to the N\'eel temperature, which magnitude is more precisely determined from temperature derivatives of both $\chi$. Both results, collected in Fig.~\ref{fig:chirhoT}(b), indicate the existence of a clear maximum at $T=(307\pm1)$~K. We assign this temperature to the N\'eel temperature of the material.
    A clear mark of magnetic phase transition is visible also in the Raman scattering spectra (see Fig.~S6 in Supplementary Materials~\cite{our-SI}), as well as in the resistivity dependence on the temperature, where a kink in the vicinity of $T_N$ was observed on the rising slope of $\rho_{xx}(T)$ (Fig.~\ref{fig:chirhoT}(a)). Interestingly, the peak in $\text{d}\rho_{xx}/\text{d}T$ occurs at the same temperature and exhibits striking similarity to the peaks in $\text{d}\chi_{\perp}/\text{d}T$ and $\text{d}\chi_{\parallel}/\text{d}T$ at $T_{\text{N}}$ (see Fig.~\ref{fig:chirhoT}(b)) together reflecting the phase transition \cite{heat1,heat2,heat3}. 

{\color{red}
    \begin{figure}[h!]
        \includegraphics[width=8.15cm]{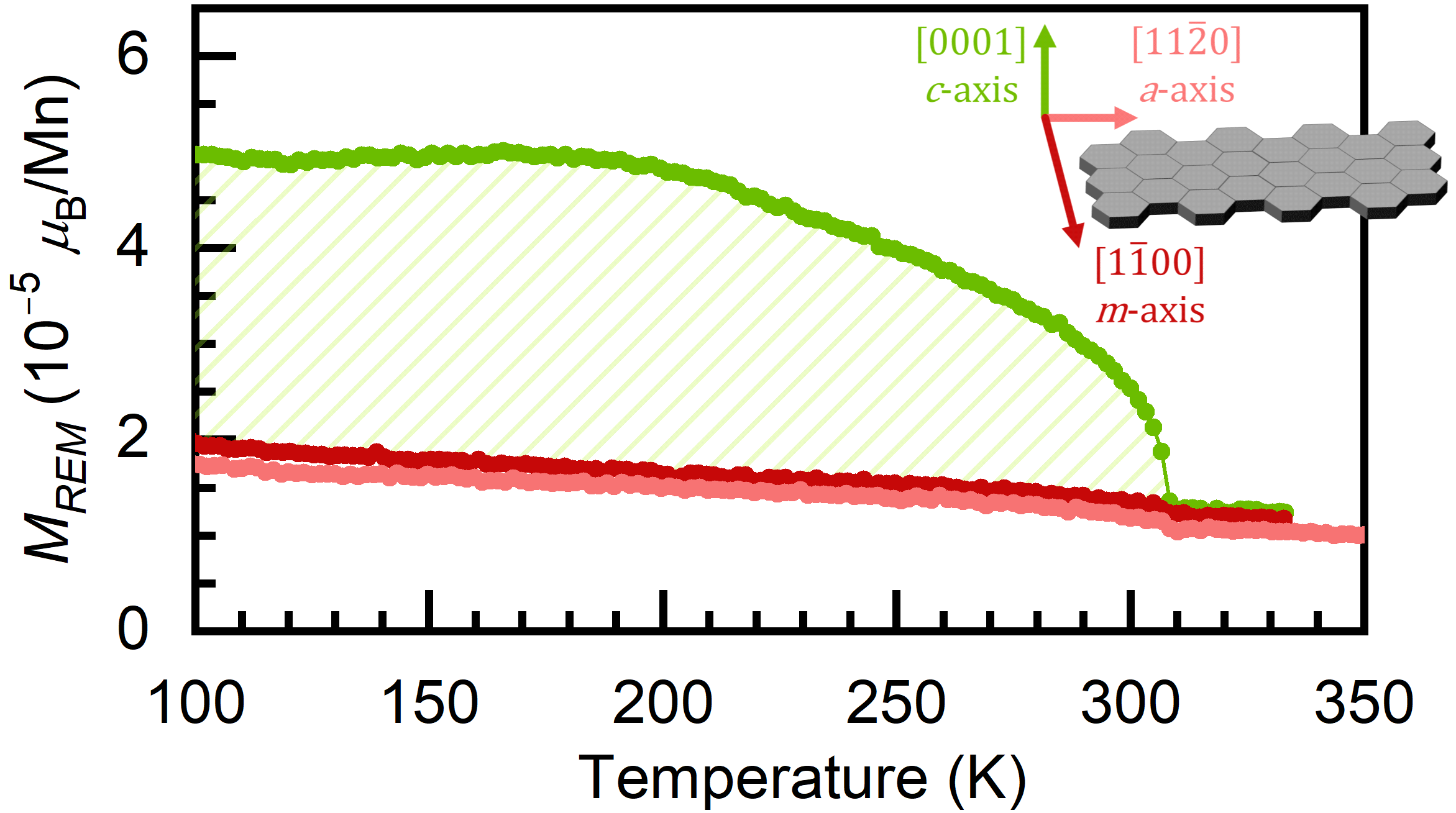}
            \caption{
            \color{black}Temperature dependence of the remnant magnetization, $M_{\text{REM}}$, acquired for all 3 major orientations of the sample. Colors of the symbols match the picture represent the experimental configuration given in the inset. The hatched area represents the temperature dependent magnitude of the WFL $M_{\text{REM}}^c$.
            \label{M_REM}\label{fig:remnantm}}
    \end{figure}
}
   
    Hex-MnTe has been regarded as a collinear AF \cite{Betancourt2023,Kriegner2016, Kriegner2017MagnAnisotrHexMnTe}, so no magnetization is expected to be seen at $H=0$. 
    First, to avoid any unintentional magnetization of the sample by a residual field in the superconducting magnet, we use the so-called „magnet reset option” to completely quench $H$. As established independently the magnitude of $\mu_0 H$ during collecting of $M _{\text{REM}}$ has been maintained below 0.01~mT. 
     
    Second, for all three orientations a non-zero magnetic moment was registered at all temperatures, including $T>T_{\text{N}}$. This feature is found in all samples studied and its origin stays currently unknown. Upon this reasoning we can treat magnitudes of both $M _{\text{REM}}^a$ and $M _{\text{REM}}^m$ as a baseline to assess true magnitudes of $M _{\text{REM}}^c$, as indicated in Fig.~\ref{fig:remnantm} by the hatched area. In following we conclude that the magnitude of $M _{\text{REM}}$ attains $3 \times 10^{-5}$~$\mu_{\text{B}}$/Mn, or the angle of canting of Mn spins from the $c$-plane towards the $c$ axis is about 0.05 deg (or 3 arc minutes). Such a very small magnitude of this $M _{\text{REM}}$ is most likely the reason it has not been revealed in the previous studies of the thin films \cite{Betancourt2023,Kriegner2017MagnAnisotrHexMnTe},  
   
    Now we note briefly the other important point that this $M _{\text{REM}}^c$ vanishes exactly at $T_{\text{N}}$. Thus it must be related to the system of antiferromagnetically coupled Mn moments and, whatever causes the Mn spins to cant, the rapidly vanishing excess $M _{\text{REM}}^c$ magnetization offers a great opportunity to directly study the properties of the N\'eel transition in this material.

    \subsection{Anomalous Hall effect}
     
    Fig.~\ref{fig:AHE} presents magnetic field dependence of the Hall resistivity $\rho_{yx}$, measured below the N\'eel temperature. The overall linear slope of the Hall resistivity has a small  hysteresis loop superimposed, which however in contrast to \cite{Betancourt2023} is of an opposite sign (see {Fig.~S3 in the Supplementary Materials~\cite{our-SI}}). If this linear Hall resistivity reflects the ordinary Hall effect, its sign points out holes as dominant charge carriers. This is consistent with the sign of thermopower voltage checked at room temperature. By subtracting the linear part, anomalous Hall component ${\rho}^{~AHE}_{yx}$ is obtained  
    (for details see Sec.~S.~III. in the Supplementary Materials~\cite{our-SI}}), where the hysteresis loop is clearly visible, see Fig.~\ref{fig:AHE}(b). The two curves come from two different pairs of Hall contacts. Both the magnitude and the shapes are very similar, pointing to the high uniformity of the sample. The AHE component stays saturated above 2~T, and at lower magnetic fields some steps are observed. Possibly they reflect locally different antiferromagnetic domain structure. It is interesting to trace temperature evolution of the Hall coefficient $R_{\text{H}}(T)$ ($R_{\text{H}}=\rho_{yx}/B$), {\color{black}the saturation value of ${\rho}_{sat}(T)$ and hysteresis width $w(T)$ of ${\rho}^{~AHE}_{yx}$.} The results are presented in Fig.~\ref{fig:all}(a)-(c).
    
    \begin{figure}[h!]
            \centering
            \includegraphics[width=8.2cm]{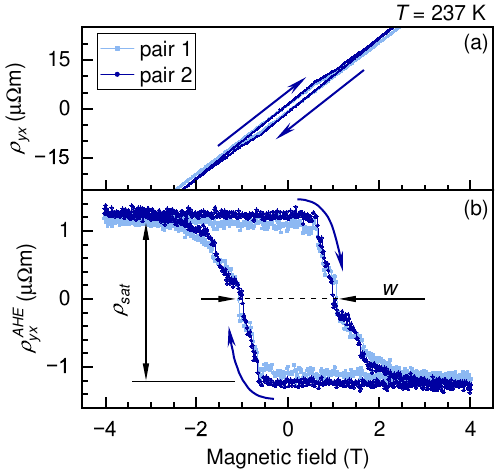}
            \caption{{\color{black}(a) Hall resistivity $\rho_{yx}$ below $T_{\text{N}}$ with clearly visible hysteresis loop. Arrows indicate direction of field sweep. (b)~The hysteresis $\rho^{AHE}_{yx}$ extracted from $\rho_{yx}$ with its characteristic parameters: width $w$ and saturation $\rho_{sat}$.}\label{fig:AHE}}
    \end{figure}
    
    \begin{figure}[h!]
            \centering
            \includegraphics[width=8.2cm]{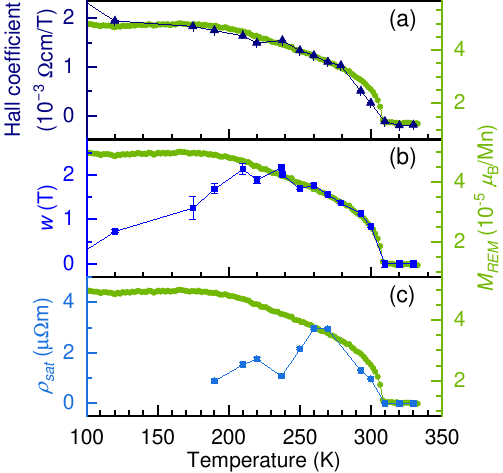}
            \caption{Comparison of the Hall effect characteristic parameters and remnant magnetization as a function of temperature: (a) Hall coefficient, (b) hysteresis width \newline and (c) saturation. \label{fig:all}}
    \end{figure}  

    The evolution of $R_{\text{H}}(T)$ shows striking resemblance to magnetic remanence dependence $M_{REM}$ on temperature, recalled for convenience on the right axis in Fig.~\ref{fig:all}(a), with a clear critical character at the N\'eel temperature. Although magnetic ions and holes are coupled systems, this Hall coefficient evolution is astonishing, revealing that the linear Hall signal cannot be exclusively related to the ordinary Hall effect dependent solely on free carrier concentration, but must be somehow influenced by the magnetic properties of the material. The results show even a change of sign of $\rho_{yx}$ close to the N\'eel temperature (see Fig.~S2 in the Supplementary Materials~\cite{our-SI}), which interpreted in a straightforward way would signify electron conductivity in the paramagnetic phase. Similar sign variation of $R_{\text{H}}$ close to the magnetic phase transition has been observed in manganese-based perovskites \cite{Li_1998} and ascribed to effects related to spin-orbit splitting. Those effects are certainly also vital here, as the top of the valence band (including curvature of the bands) is strongly influenced by spin-orbit interaction~\cite{pfjr2023} but microscopic mechanism could be very different in MnTe (we elaborate on this in the Supplementary Materials~\cite{our-SI}).
   
    The temperature impacts also the shape of the hysteresis loop. First of all, the hysteresis vanishes (stays below our detection limits) above 307 K (see Fig.~\ref{fig:all}(b)-(c)), clearly indicating relation with the magnetically ordered system of Mn ions. The $\rho_{sat}(T)$ and $w(T)$ initially follow $M_{REM}$, however in a more limited temperature range. Similar observation have just been made by R.~D.~Gonzalez~Betancourt~et~al.~\cite{Betancourt2023} for thin MnTe films and by Y.~Liang~et~al.~\cite{Liang2023} for metal/antiferromagnetic insulator heterostructures, where a model of antiferromagnetic topological textures was proposed with colinear AF deflecting due to both AF exchange coupling and Dzyaloshinskii–Moriya interactions (DMI). Although the studied system here is different, same phenomena occur as well, with the DMI-like interactions playing an important role.

    We argue below that $M_{REM}$ is in principle not necessary for the AHE to occur,
    were we somehow able to prepare the system in a single-domain state. However, 
    if the $\vec{N}$ and $-\vec{N}$ states (in other words, $\updn$ and $\dnup$)
    have the same energy and occur with equal probability then the AHE will average
    out to zero as demonstrated in Fig.~1(e) of Ref.~\cite{Betancourt2023}. Combined
    effect of DMI and external magnetic field
    can lift the degeneracy of $\updn$ and $\dnup$ in energy, thereby allowing for 
    AHE that does not average out to zero, but symmetry of NiAs does not allow for DMI.
    Finite value of $M_{REM}$, as found experimentally suggests that a more complicated interaction between Mn magnetic moments is
    present instead, that lifts the degeneracy of $\updn$ and $\dnup$ when magnetic
    field along $c$-axis is applied and we elaborate on this argument in the subsequent
    section.


    
\section{Discussion of the origin of AHE} 

The origin of the anomalous Hall effect is complex, depending on the studied system \cite{Nagaosa2010, Jungwirth2002, StredaVyborny2023}. Generally, the diverse mechanisms behind it are classified into intrinsic ones, related to the symmetry of the band structure and extrinsic, related to the scattering mechanism of charge carriers, all leading to an additional correction $\rho^{AHE}$ to Hall component of the resistivity tensor 
\begin{equation}
    \rho_{\text{H}}=R_{\text{H}}\cdot B+\rho^{AHE}, \label{eqn:AHE}
\end{equation}
whereas traditionally, $\rho^{AHE}\propto m$ was assumed --- based on phenomenology for most ferromagnets~\cite{Nagaosa2006} --- and the excitement about AHE in antiferromagnets has been coming from the realisation that non-zero $\rho^{AHE}$ can indeed occur even if the magnetization $m$ vanishes.
  


The observed coexistence of finite magnetic remanence $M_{REM}$ and AHE 
may lead to the attempt of describing the AHE as originating solely from $M_{REM}$. We believe however that the situation is more complex. There are numerous systems in which in spite of $M_{REM}=0$ still the AHE is predicted (see e.g.~\cite{Chen_AHE_2014}) and explained by wavefunction symmetry considerations and non-vanishing Berry curvature. 
Below we show that for some N\'eel vector directions Berry curvature of the MnTe valence band stays finite, thus contributing to the AHE~\cite{Nagaosa2010}. Then we switch to modelling of impact of this weak magnetic signal within a macrospin model, where finite magnetic moment is evoked by the 
interaction of the Dzyaloshinskii–Moriya type and it can reproduce well the hysteretic behavior of AHE.



\subsection{Berry curvature}

{\color{violet} 
}

\begin{figure}
            ~\\~\\ 
            \centering
            \includegraphics[width=8cm]{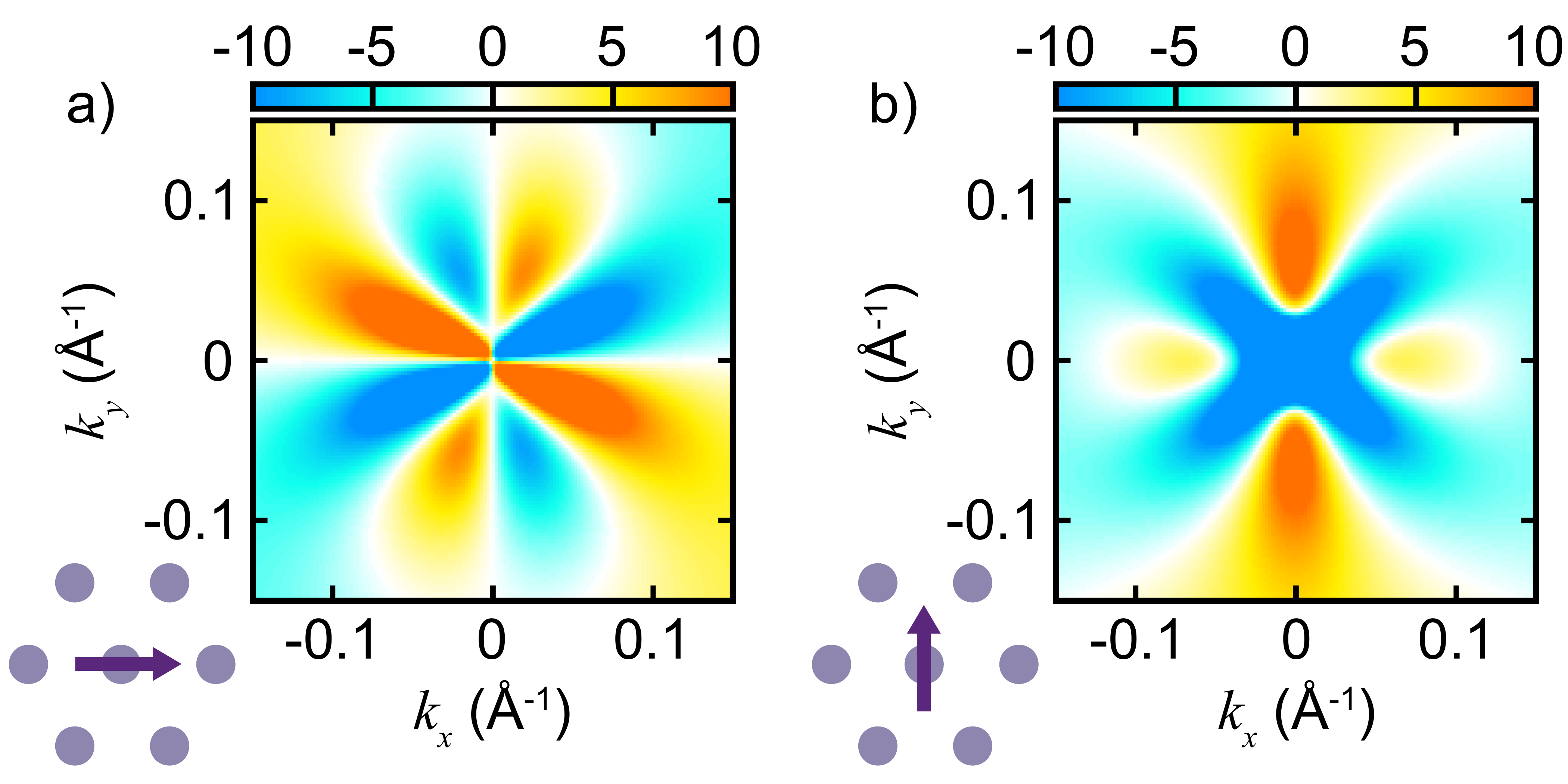}
            \caption{Berry curvature (in $\AA^2$) of the topmost valence band in the
            vicinity of its maximum (section $k_z=\text{const.}$ is shown, $k_{x,y}$ in $\AA^{-1}$; see {Sec.~S.~V. in Supplementary Materials~\cite{our-SI}} for further details). {\color{black} Position of the N\'eel vector is indicated at the foot of the image}
            \label{fig-07}} 
    \end{figure}

Symmetry arguments~\cite{Betancourt2023} regarding the existence of AHE are valid irrespective of its detailed microscopic mechanism. If we assume that the origin of
this effect in MnTe is intrinsic, we can attain additional understanding regarding
the on/off switching of AHE (depending on N\'eel vector direction) by
analysing the Berry curvature $\Omega_z(\vec{k})$. In Fig.~\ref{fig-07},
we consider a section of the crystal-momentum space (i.e. $\vec{k}$ space) 
in a plane perpendicular to $\Gamma$A line so that the situation around the 
valence band top is shown (this choice is explained in {Supplementary Materials~\cite{our-SI}, Sec.~S.~V.}).
Unlike in ferromagnets where the intrinsic AHE is believed to originate from
hot spots of Berry curvature~\cite{Fang2003}, its structure in MnTe is richer,
comprising minima and maxima (such character has recently been analysed in terms
of {\em multipoles}~\cite{Zhang:2023_a} as opposed to the monopole character of 
the aforementioned hot spots) and their balance depends on the orientation of the
N\'eel vector. 

We now turn our attention to this feature of the BC.
On the left panel of Fig.~\ref{fig-07}, the integral of $\Omega_z^{k_z}(k_x,k_y)$
is zero: even if the BC is non-trivial also for this direction of the N\'eel 
vector, Fermi sea integral is still zero and the AHE is forbidden. For other directions of
the N\'eel vector (except $120^\circ$ rotated cases), this symmetry is broken and
the BC has non-zero integral. As an example we show $\vec{N}\parallel y$ on the right panel.
We conclude that while the BC in ferromagnets has a monopole character in reciprocal
space~\cite{Nagaosa2006}  
its structure in AF is less trivial. Since collinear magnetic order is simpler than
e.g. kagome-lattice AF systems (such as Mn$_3$Sn~\cite{Nakatsuji2015}), the symmetry
of BC plots can be straightforwardly analysed: from the case of N\'eel vector parallel
to $c$-axis (not shown) where three-fold symmetry is preserved to $\vec{N}\parallel x$
where symmetry is lowered but integrated BC remains zero and finally to cases as shown 
in Fig.~\ref{fig-07}(b) where the AHE becomes allowed.

\subsection{Modelling of AHE hysteresis loop}

To demonstrate how external field applied along the $c$-axis lifts the degeneracy of energies of $\updn$ and $\dnup$ domains, 
we now attempt to describe the behavior of a weak non-zero magnetization found
in experimental data of Fig.~\ref{fig:remnantm}.
Such situation can arise due to the DMI that is known to be responsible for weak magnetization in many systems such as hematite \cite{Zhang2019}. 
Symmetry of NiAs-type MnTe allows for a higher-order interaction~\cite{Wasscher:1965_a}
that shares the property of creating an uncompensated net magnetic moment with DMI.
In its basic form $\mathbf{D}_{12}(\mathbf{M}_1 \times \mathbf{M}_2)$, the DMI is forbidden in NiAs-type structure but we use a more general approach as follows. We construct a 2D macrospin model in which we consider two antiferromagnetic moments coupled by exchange interaction $J$, in the external magnetic field $\mathbf{B}$, uniaxial anisotropy $K_u$ and 
an interaction related to the spin-orbit parametrised by $D_{12y}$.
The set of coordinates is chosen so that the magnetic moments lie in $xz$-plane, which corresponds to a cross section through a magnetic-easy $xy$-plane of MnTe and the magnetic field is applied out of plane. The energy expression has the following form:
\begin{widetext}
\begin{equation}
\varepsilon=J\mathbf{M}_1\cdot\mathbf{M}_2-\mathbf{B}\cdot(\mathbf{M}_1+\mathbf{M}_2)+K_uV\left(\sin^2{\alpha_1}+\sin^2{\alpha_2}\right)+D_{12y}f(\phi,\gamma) \label{eqn:dmi}
\end{equation}
\end{widetext}
where $\alpha_{1,2}$ describes the angle between a corresponding magnetic moment and a common crystalline direction within the $xz$-plane . 
The first three terms correspond to the usual~\cite{Volny:2020_a}
Stoner-Wohlfarth model of an antiferromagnet with two magnetic sublattices. If
$\mathbf{D}_{12}||\hat{y}$ were allowed by symmetry 
then $f$ would only depend on the canting angle $\gamma=\alpha_2=\pi-\alpha_1$. In our case,
the basic form of Dzyaloshinskii-Moriya is replaced by a more complicated functional
dependence~\cite{Wasscher:1965_a} (including the azimuthal angle $\phi$) but when the moments are confined to $xz$-plane,
and $|\mathbf{M}_1|=|\mathbf{M}_2|=M_0$,
only $f=M_0^2\sin(2\gamma)$ remains as the canting angle $\gamma$ is the
same for both sublattices.


%
The energy can be explicitly expressed as function of these two angles $E(\alpha_1,\alpha_2)$ and consequently its minimization brings information about the expected equilibrium state of the system. It is convenient to divide the Equation~\ref{eqn:dmi} by the sublattice saturation magnetization $M_0$ as the resulting $E=\varepsilon/M_0$ is equivalent to $\varepsilon$ from the minimization point of view. We perform numerical calculation of the energy landscape for all possible angles $\alpha_{1,2} \in [0,2\pi]$ and chosen set of parameters expressed in Teslas: $B_J,B_K,B_D$, where $B_J=J M_0$ is exchange field, 
{\color{black}$B_K = K_uV/M_0$}
is anisotropy field and $B_D=D_{12y}M_0$ is the DMI-like field and $V$ is the volume of the system. We notice that for $B=0$ we obtain two equivalent energy minima both for zero and non-zero $B_D$. Moreover, for the chosen set of parameters, it is enough to minimize $\epsilon$ with respect to one of the $\alpha$, because $\alpha_2$ is dependent on $\alpha_1$ and can be unequivocally extracted: they are coupled via the exchange interaction which is very strong compared to the external magnetic field.

\begin{figure}[h!]
    \centering
    \includegraphics[width=8.5cm]{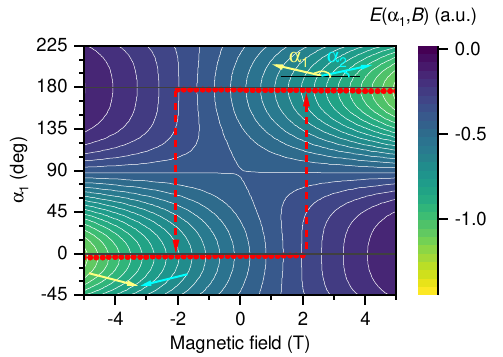}
    \caption{\label{fig:ELH} {Energy landscape as a function of external magnetic field applied along $z$ axis and the orientation of one magnetic moment with respect to the easy-plane described by $\alpha_1$ calculated for $B_J=100$~T, $B_K=0.04$~T and $B_D=10$~T. The red points indicate the position of energy minima.}}
\end{figure}

Energy landscape may exhibit multiple minima for small values of external magnetic field. An example of such case is presented in Fig.~\ref{fig:ELH}, where the energy colour map is shown as a function of magnetic field along $z-$axis and $\alpha_1$. It can be easily seen that while for $B_z=0$ two states producing magnetization with different signs along $z$ are equivalent and although strong magnetic field favours one of them, there is still a region of moderate fields in which the magnetization can have  two competitive positions due to the local energy minimum (minima are marked by red dots in Fig.~\ref{fig:ELH}). This observation can reproduce the appearance of hysteresis in experiments while sweeping the magnetic field as shown by the red dots. The red dashed arrows indicate the magnetic field at which an abrupt N\'eel vector reorientation occurs.

The disappearance of the local energy minimum can be clearly seen in {Fig.~S4, presented in the Supplementary Materials~\cite{our-SI}}, where only few chosen energy curves for different fields are presented. Non-zero external magnetic field changes the number of available energy minima. Upon increasing the magnitude of the magnetic field local energy minimum around $\alpha_2=0$ disappears. This is expected to cause the reorientation of the magnetic moments since the $\alpha_2$ will change by nearly $180^\circ$ to reach the new energy minimum. As a result, the net magnetic moment will reorient to the direction parallel to the magnetic field if it was antiparallel originally. This scenario provides hysteretic behavior of magnetization.
    

We notice that the hysteresis can be reproduced for certain range of parameters, and the width of the hysteresis depends on $B_D$ and $B_K$ ({Fig.~S5 in the Supplementary Materials~\cite{our-SI}}). 
Typically, values of the exchange fields are orders of magnitudes higher the anisotropy fields \cite{Kittel1951, Keffer1952, Kimel2004}. We keep $B_J=100\,\text{T}$, which is a common general assumption \cite{Kittel1951, Keffer1952, Kimel2004}, very reasonable agreement with \cite{Szuszkiewicz2006} and the same order of magnitude as \cite{Kriegner2016}. The anisotropy field values used for the simulations are below $0.1\,\text{T}$ (common limit in the literature is $<1\,$T \cite{Kittel1951, Keffer1952, Kimel2004}, see also \cite{Lebrun2019}). Then, for such set of parameters we notice that $B_D$ has to reach values at the level of single Teslas for the onset of hysteretic behavior of magnetic orientation in the external magnetic field. If compared to the values of the DMI fields, we notice that while interfacial DMI tend to have significantly lower values \cite{Kuepferling2023}, the DMI field in hematite, which is a canted antiferromagnet, was determined to be $2.72\,$T \cite{Lebrun2019}, of the same order of magnitude as in our case. When $B_D$ is too small compared to $B_J$, the two energy minima persist for all tested magnetic fields similarly to the case of fully compensated antiferromagnet where the two opposite in-plane states of the N\'eel vector remain degenerate in the out of plane magnetic field. When $B_D$ is large, magnetic moments may still reorient close to zero field but the hysteresis may become invisible. Additionally {(as shown in Fig.~S5 in the Supplementary Materials~\cite{our-SI})} increasing anisotropy may enhance the hysteresis width which may explain the temperature evolution of the observed hysteresis (Fig.~\ref{fig:all}). We also emphasise that our model reveals an interesting mechanism in which the N\'eel vector rotates by $180\,^\circ$ upon sweeping the magnetic field magnitude in perpendicular direction.
 

\section{Summary}

We report on the observation of a hysteretic behavior of transverse resistivity in the magnetic field that can be ascribed to anomalous Hall effect as well as detection of small but measurable magnetic moment in bulk, antiferromagnetic hexagonal MnTe. The AHE hysteresis width, magnitude and the remnant magnetization - exhibit similar dependencies on temperature in the high temperature range, i.e. they monotonously decrease with temperature and vanish above $T_{\text{N}}$, which is carefully determined by multiple, independent experimental methods. This clear correlation between the magnetization and transport data paved the way to construct a macrospin model with additional, phenomenological energy term inspired by Dzyaloshinskii–Moriya interaction, in which the uncompensated magnetic moment is found to be responsible for the hysteretic behavior, within a ceratain range of material parameters. The direct discrimination between different magnetic and non-magnetic effects to the magnetotransport data and detailed discussion of their magnitudes remains challenging and therefore the existence of other mechanisms giving rise to AHE in our samples cannot be unequivocally ruled out.

%
%

%

\begin{acknowledgments}
    This work was partially supported by the Polish National Centre for Research and Development through grant No. TECHMATSTRATEG1/346720/8/NCBR/2017 and by the National Science Centre, Poland under Grant 2021/40/C/ST3/00168. Czech Science Foundation
    (GA\v CR) provided support under grant 22-21974S. JS acknowledges a support within "New Ideas 2B in POB II” IDUB project financed by the University of Warsaw.
    $MG$-$B$ and $JS$ express their gratitude to prof.~W.~Natorf. We 
    are indebted to Prof.~U.~Zuelicke for stimulating discussions and
    KV thanks O.~Sedl\'a\v cek for assistance in numerical analysis of Berry curvature. 
\end{acknowledgments}



\section*{Data Availability}
    The data that support the findings of this study are available from the corresponding author upon reasonable request.
   
\nocite{*}
\bibliography{0-pub_main}








\end{document}